\documentclass[showpacs,showkeys,superscriptaddress,preprintnumbers,amsmath,amssymb,longbibliography,notitlepage]{revtex4-1}

\usepackage{epsf}
\usepackage{graphicx}
\usepackage{dcolumn}
\usepackage{bm}
\usepackage{subfigure}
\usepackage{color}
\usepackage{amsthm}
\usepackage{amssymb}
\usepackage{amsmath}
\usepackage{pb-diagram}

\usepackage{hyperref}

\newcommand{\setof}[1]{\left\{ {#1}\right\}}

\def\setof#1{\left\{{#1}\right\}}

\definecolor{gray85}{gray}{0.85} 
\definecolor{gray8}{gray}{0.8} 
\definecolor{gray7}{gray}{0.7} 
\definecolor{gray6}{gray}{0.6} 
\definecolor{gray5}{gray}{0.5} 
\definecolor{gray4}{gray}{0.4} 
\definecolor{gray35}{gray}{0.35} 

\newcommand{\pd}{{\mathsf{PD}}}

\begin{document}

\title{Granular Response to Impact: Topology of the Force Networks}

\author{T. Takahashi}
\affiliation{Department of Mathematical Sciences, New Jersey Institute of Technology, Newark, NJ 07102}

\author{Abram H. Clark}
\affiliation{Department of Physics, Naval Postgraduate School, Monterey, CA 93943, USA}

\author{T. Majmudar}
\affiliation{Department of Mathematics, New York University, NYC, NY 10012}

\author{L. Kondic}
\affiliation{Department of Mathematical Sciences, New Jersey Institute of Technology, Newark, New Jersey 07102}

\begin{abstract}
Impact of an intruder on granular matter leads to formation of mesoscopic force networks seen particularly clearly in the recent experiments carried out with photoelastic particles (Clark \textit{et al.}, Phys. Rev. Lett., {\bf 114} 144502 (2015). These force networks are characterized by complex structure and evolve on fast time scales. While it is known that total photoelastic activity in the granular system is correlated with the acceleration of the intruder, it is not known how the structure of the force network evolves during impact, and if there is a dominant features in the networks that can be used to describe intruder's dynamics. Here, we use topological tools, in particular persistent homology, to describe these features. Persistent homology allows quantification of both structure and time evolution of the resulting force networks. We find that there is a clear correlation of the intruder's dynamics and some of the topological measures implemented. This finding allows us to discuss which properties of the force networks are most important when attempting to describe intruder's dynamics. Regarding temporal evolution of the networks, we are able to define the upper bound on the relevant time scale on which the networks evolve. 
\end{abstract}
\date{\today}


\maketitle

\section{Introduction}

When a high-speed intruder strikes a granular material, its momentum is carried away and dissipated by the grains. This process is important in a wide variety of natural and man-made settings, including astrophysics~\cite{Moore1980,Colwell1999}, rugged-terrain robotics~\cite{raibert2008bigdog,li2013terradynamics}, and ballistics~\cite{stoffler1975}. Previous experimental studies~\cite{poncelet,Allen1957,Forrestal1992,durian_pre03,Uehara2003,goldman_prl04,durian_pre05_71,durian_nat07,goldman_pre08,durian_prl08,umbanhowar_pre10, brzinski2013,nordstrom2014,Chen2014,jerome2016}, have developed macroscopic descriptions involving, for example, the dynamics of the intruder~\cite{Allen1957,durian_nat07,goldman_pre08}, the size of the impact crater~\cite{Uehara2003,durian_pre05_71}, collective dynamics of multiple intruders~\cite{durian_prl08}, the influence of bed preparation~\cite{umbanhowar_pre10}, or the influence of interstitial fluid~\cite{nordstrom2014,jerome2016}. However, relating macroscopic behavior to physical processes at the grain scale or at intermediate length scales can be quite difficult, primarily because measuring forces and dynamics inside the granular material is not possible in most experimental realizations.

Recent experiments on granular impact using photoelastic disks~\cite{wakabayashi60,takada76,baxter97,daniels2017photoelastic} and high-speed video (frame rates of 10-50 kHz) have provided some insight on the complex nature of grain-grain force transmission within the granular material during initial impact~\cite{clark_prl15} and penetration~\cite{clark_prl12}. This allows visualization of the force networks that form during impact, and these forces can then be connected to the intruder dynamics~\cite{clark_prl12,clark_pre14,clark_epl13} or the motion of grains~\cite{clark_pre16}. However, a complete description of the relationship between the intruder motion and the space- and time-dependent granular forces is still lacking, partially because the images from high-speed videos lack the spatial resolution required to quantitatively measure vector forces~\cite{majmudar05a,majmudar07a} between grains. Quantitative photoelastic measurements are thus limited to, for example, the total photoelastic intensity in a certain region of the image, which gives an estimate of the pressure in that region.

In recent years, significant progress has been made toward an understanding of force networks using tools such as statistical methods~\cite{peters05,tordesillas_pre10,tordesillas_bob_pre12}, network analysis~\cite{daniels_pre12, herrera_pre11,walker_pre12}, and application of topological methods~\cite{arevalo_pre10, arevalo_pre13, epl12,pre13,physicaD14,pre14}. The cited works mostly focus on the analysis of the data resulting from simulations, and there have not been attempts so far to analyze the properties of force networks that emerge during dynamic experimental processes evolving on fast time scales. 

In this work, we will analyze the granular force networks that form during impacts into photoelastic disks using the tools of  persistent homology. Persistent homology involves a topological characterization of a scalar field, in this case the brightness of a  given photoelastic image from a high-speed movie of a granular impact event. The resulting topological measures can give detailed information about the granular response at a spatial scale that is larger than a grain but small compared to the size of the granular assembly. By analyzing a time series of photoelastic images, we show that persistence homology can provide more detailed connection between the granular force response and the intruder dynamics than can be provided by simply measuring the total photoelastic response. To our knowledge, this is the first attempt to use topology-based methods to analyze the results of physical experiments on granular materials involving fast dynamics.

The rest of this manuscript is organized as follows. In Section~\ref{sec:techniques} we discuss the techniques used to analyze the results, including experimental techniques, image processing, and a brief description of the used topology-based measures. In Section~\ref{sec:results} we discuss the main results, including both structural and time-dependent properties of the force networks. Section~\ref{sec:conclusions} gives conclusions and outlook.

\section{Techniques}
\label{sec:techniques}
\subsection{Experimental Techniques}

We perform topological analysis on images from high speed impacts into a collection of photoelastic disks (3~mm thick), which are confined between two Plexiglas sheets (0.91~m $\times$ 1.22~m $\times$ 1.25~cm) separated by a thin gap (3.3~mm). The experimental apparatus is identical to that used in previous experiments~\cite{clark_prl12,clark_epl13,clark_pre14,clark_prl15}. Circular intruders machined from bronze sheet (bulk density of 8.91~g/cm$^3$ and thickness of 0.23~cm) impact into disks of diameters $D$ of 6.35~cm, 12.7~cm, and 20.32~cm. We drop these intruders from varying heights through a shaft connected to the top of the apparatus, yielding an impact velocity $v_0\leq 6.6$~m/s. We record results with a Photron FASTCAM SA5 at frame rates of up to 25,000 frames per second (fps). To measure $v_0$, we track the intruder and take a numerical derivative as in~\cite{clark_prl12,clark_epl13,clark_pre14,clark_prl15}.

Photoelastic particles are cut from polyurethane sheet from Precision Urethane into disks of 6 mm and 9 mm diameter. We use two different sets of particles with distinct stiffnesses: Shore 60A (bulk elastic modulus of roughly 1-10 MPa) and Shore 80A (bulk elastic modulus of roughly 10-100 MPa). We give a more comprehensive treatment of the physical properties of these particles, including a relation of force as a function of compression, in Ref.~\cite{clark_prl15}. Following~\cite{clark_prl15}, we refer to Shore 60A particles as soft and Shore 80A particles as medium stiffness.

\subsection{Image Processing}

\begin{figure}
 \includegraphics[width=1.0\columnwidth]{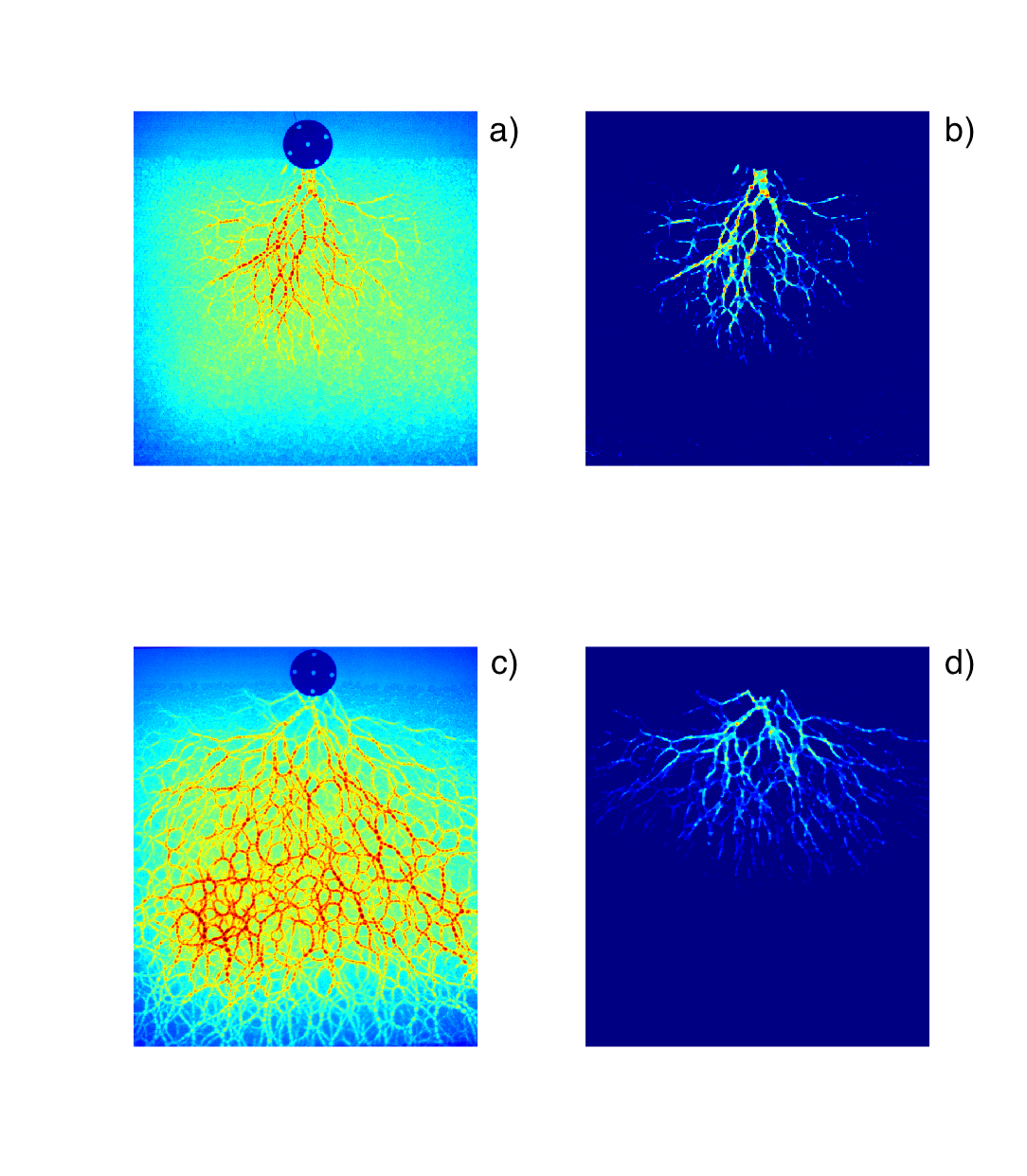}
\caption{Example of image processing for medium $6.35-4.5$ (a,b) and soft $6.35-1.3$ (c,d) particles. Here, (a,c) are raw images (with a false color scheme added) and (b,d) show the post-processed images (see text for details on image processing). The naming convention is such that the first number specifies radius of the intruder in cm, and the second one the impact speed in m/s. The images are shown at the times $0.276$ ms (medium) and $16$ ms (soft) after impact. We will refer to medium $6.35-4.5$ experiment as the {\it reference} one for the remainder of the manuscript.
}
\label{fig:processing}
\end{figure}

Figure~\ref{fig:processing} shows two examples of experimental images, for impact on medium (a,b) and soft (c,d) particles. We first remove noise from the initial images, (a,c) in  Fig.~\ref{fig:processing}, by spatial homogenization to account for the inconsistent lighting as well as by applying a notch filter to remove the flickering of the AC light source at 120~Hz. We then perform image subtraction between each frame and a reference frame. The result is an 8-bit grayscale image that shows negligible temporal fluctuations or spatial inconsistencies in the lighting. We note that, for the soft particles, pre-existing force chains from gravity are visible before impact, as seen in Fig.~\ref{fig:processing}c), and these are subtracted from the final images. After noise removal, we use built-in MATLAB functions to slighly dilate and erode the images, to make sure gaps between neighboring grains in a force chain are connected. The final images are as shown in Fig.~\ref{fig:processing}(b,d). The image processing we show is performed on the entire image. 

Figure~\ref{fig:processing}(a,b) corresponds to medium stiffness particles, for which the speed of sound is roughly $90$ m/s~\cite{clark_prl15}.  These images are recorded at 
$25,000$ fps. For soft particles shown in Fig.~\ref{fig:processing}(c,d), the estimated speed of sound is $30$ m/s~\cite{clark_prl15}, and the frame rate here is $10,000$ fps.  
The speed of sounds divided by the frame rate gives the typical distance of information propagation between two consecutive images of roughly $3$ mm, which is 
about half of the smallest grain diameter. Thus, we record images sufficiently fast to resolve grain-grain force transmission during impact, although our results below suggest the possibility of evolution on even faster time scales.

\subsection{Topological Measures}
\label{sec:topology}

In the context of granular matter, persistence homology computations have so far been reported for the data from  numerical simulations. The results presented here is the first attempt to extract useful information using persistent homology from experimental data. In our case, the data available, consisting of the images shown in the previous section, are by necessity incomplete, since the impact is a fast process involving large number of particles. Therefore the information about individual contact forces cannot be obtained. These images cannot be used to distinguish the forces at each contact, therefore detailed information about the interparticle forces is not available. Thus, the question we address is whether useful information can be extracted from such data using persistent homology computations. If so, then the tools we discuss here could be applied to a broad range of experimental granular systems involving fast dynamics.

The details regarding applications of persistent homology to analysis of force networks in granular matter could be found in previous works~\cite{physicaD14,pre14}, and here we provide a brief overview. Before going into the description, it is important to point out one significant difference between the approach taken by the persistence homology compared to more classical approaches, that typically consider the interparticle forces as a function
of (often arbitrarily chosen) force threshold. Instead of using a threshold, persistent homology is able to treat the information about the forces on all levels at once. Therefore, separation into weak and strong force networks, for example, is not necessary, although it can be carried out if so desired. For the data considered in this paper, we will not be even considering forces on the level of contacts, as discussed above, but on the level of image brightness.

Each experimental image could be considered as a rectangular array of pixels, with each pixel specified by its brightness value in the range $\theta \in [0:255]$. To analyze such an image, we carry out {\it filtration}, meaning that (in the case considered) we consider the pixels with the brightness {\it above} specified threshold as white, with the rest being black (superlevel filtering). Clearly, if we choose very high threshold, only few pixels will be white. Then, as we lower the threshold level, more and more pixels will become white. In simple terms, persistent homology keeps track and quantifies the {\it connectivity} of the network that forms by the white pixels. For detailed discussion of dealing with pixelized data in the context of particulate systems we refer the reader to~\cite{physicaD14}; a less technical description could be found in~\cite{pre14}. We note however, that the listed references focus on extracting force networks, since the data (either experimental examples or simulation results) included sufficient information to be able to extract the forces at each particle contact. This is not the case here, so we focuses on the direct analysis of the pixelized images, without attempting to formulate well-defined force networks. The codes used for computation of persistence are available online~\cite{codes}.

By the filtering process described above, each image could be associated with the corresponding {\it persistence diagrams}, PDs, that quantify changes of the image as the filtering threshold, $\theta$, is modified. The $\beta_0 (\beta_1)$ PDs, PD$_0$ and PD$_1$, respectively, encode the appearance and disappearance of connected components and loops, by keeping track of the values of $\theta$ at which a component and/or loop appear or disappear. To help interpretation, one could think of the `connected components' as `force chains' (for large values of $\theta$), although the concept itself is more general (and much more precisely defined)  than rather vague `force chain' one. The PD$_1$ keeps track of appearance and disappearance of loops, defined by the requirement that all pixels that form a loop are brighter than a specified value of $\theta$. Connected components disappear when they merge (at some lower value of $\theta$);  the loops  disappear when they get filled up with the pixels that are at least as bright as the specified threshold. 

\begin{figure}
 \includegraphics[width=1.0\columnwidth]{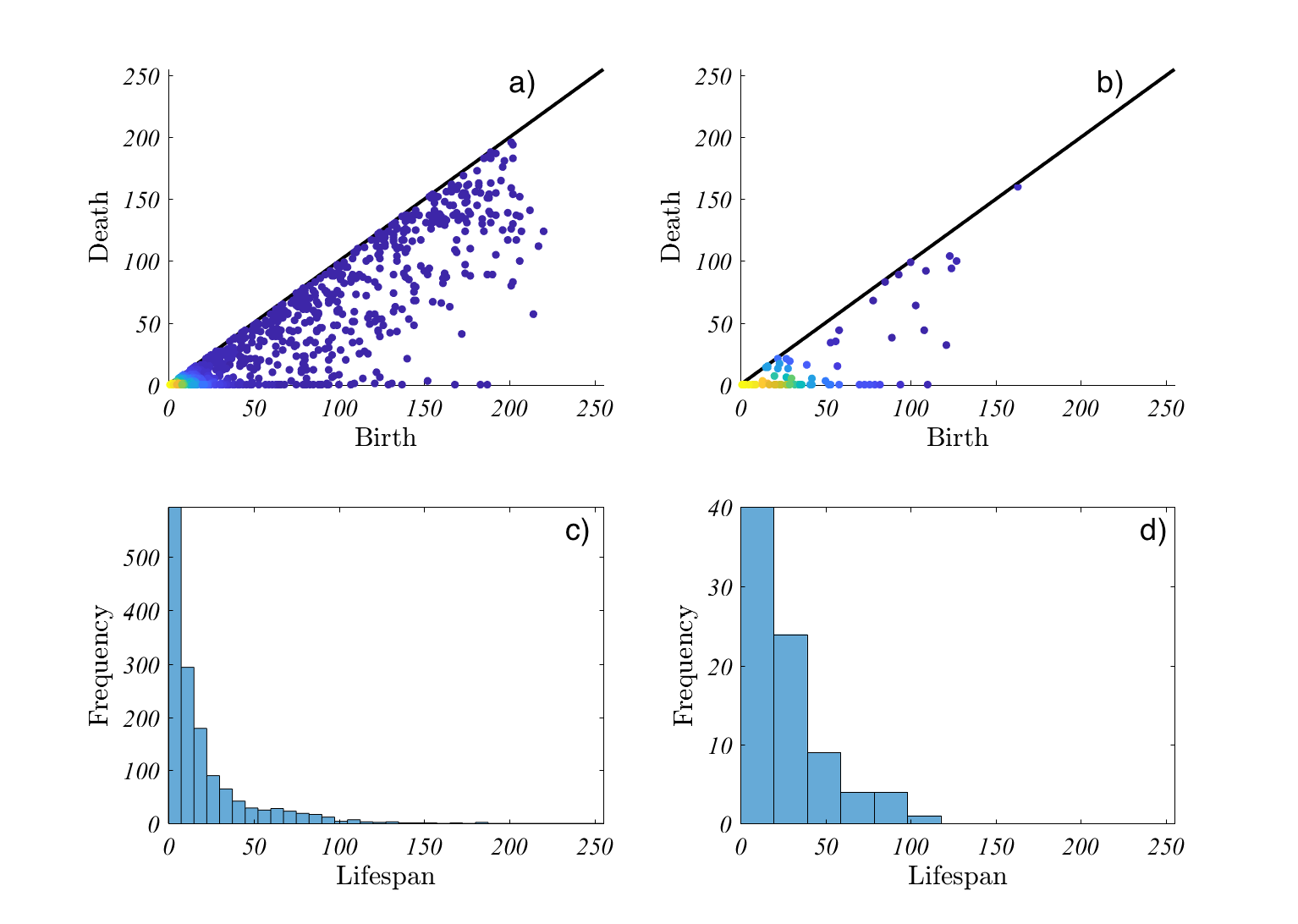}
\caption{Persistence diagrams (PDs) for the reference experiment (medium $6.25-4.5$) (same time as shown in Fig.~\ref{fig:processing}). 
The parts (a) and (b) show the PDs for components and loops, PD$_0$, and PD$_1$, respectively. The parts (c) and (d) show the corresponding 
lifespans. The $x$-axis range correspond to brightness, covering the full range $[0:255]$. Note la:/Urge number of generators with very small lifespans, 
showing that there are many points in the diagrams that are near the diagonals and represent essentially noise. 
Supplementary animation~\cite{sup_anim_diagrams} show the PDs and lifespans for all considered images of the reference experiment. 
 }
\label{fig:diagram}
\end{figure}

Figure~\ref{fig:diagram} shows an example of PDs for one image/frame from the reference experiment. On the horizontal axis we plot birth (appearance) of a component/loop, and at the vertical axis we show the disappearance (death) coordinate. Part (a) shows the diagram for the components, PD$_0$, and the part (b) shows the diagram for the loops, PD$_1$. Since components/loops are born before they die, all the points (generators) are below the diagonal. To guide the reader in interpreting these diagrams, we provide a few brief remarks:
\begin{itemize}
\item 
Note large number of generators close to the diagonal. These are generators that persist just for a small range of brightness values and could be considered as noise. 
\item
The dominant features (such as strong `force chains') are described by the generators that are further away from the diagonal, and which are 
born at high values of $\theta$. 
\item Note that the birth values of PD$_1$ are lower than that of PD$_0$. This is due to the fact that the birth of a loop corresponds to the 
lowest magnitude of the pixels that form the loop. 
\end{itemize}

These PDs are essentially point clouds that describe (in a simplified manner) complex network structure. They are still, however, rather complex
since they include the information about connectivity of the networks over {\it all} brightness levels. Therefore, one needs some 
approach for their analysis and quantification. While there are different quantities that could be used for this purpose, in the present 
context we find it useful to consider the following simple measures:
\begin{itemize}
\item 
{\it Lifespan:} Consider a generator that got born on the level $\theta  = b$, and died on the level $\theta = d$. The lifespan is then defined as $b - d$. 
The lifespans for the PDs shown in Fig.~\ref{fig:diagram} (a) and (b) are given in the parts (c) and (d), respectively, of the same figure. We note that 
small lifespans occur very often, saying that there are many generators very close to the diagonals. In our data analysis presented in the next section, 
we find it convenient in some cases to remove such small lifespans from consideration, since the corresponding generators could 
be thought of as a consequence of noise in the experimental images. 
\item
{\it Total persistence} (TP)  is defined as the sum of all lifespans, TP(PD) $= \sum_{(b,d) \in {\rm PD}}(b - d)$. 
TP allows to describe a whole PD by a single number, of course at the cost of a significant information
loss. If we think of the original image as a landscape (with the altitude corresponding to pixel brightness)  then the TP essentially tells us how flat this landscape is:
larger TP corresponds to the landscape that is more mountain-like, with lots of hills and valleys, while small TP corresponds to the landscape without too many 
features. Since we have PD$_0$ and PD$_1$, we define total persistence for the components and the loops,  TP$_0$ and TP$_1$, respectively. 
Another related measure is the total number of generators, TC, again for the components, TC$_0$, and loops, TC$_1$. 
\item {\it Betti numbers:} Betti numbers simply count the number of features, components, $\beta_0$, or loops, $\beta_1$, at a specified threshold level, $\theta$. 
These quantities could be obtained by simple counting, persistence homology is not required for their definition or computation. However they could be also obtained from 
the PDs by essentially summing the number of generators such that their birth coordinate is larger, and death coordinate smaller, than a considered
threshold, $\theta$. It should be noted that Betti numbers are function of $\theta$, and therefore contain much less information than PDs themselves. 
\end{itemize}

It should be noted that it can be proven that PDs are stable with respect to noise (small changes of input data produce small changes in corresponding PDs); however, 
the same cannot be shown for Betti numbers~\cite{physicaD14}. This property of PDs and Betti numbers should be remembered in particular when dealing with potentially 
noisy data, as the ones considered in this paper.

The final quantity that we discuss in this overview of topological measures is the concept of distance between the PDs. Definition of such a quantity is possible since 
the space in which PDs live is a metric space and therefore the concept of distance can be defined. The metrics that we consider and define here is based on 
the entire diagram, i.e.\ we compare two diagrams by comparing all points in each diagram. Note that this comparison does not involve thresholding: this measure
compares the images on all brightness levels. 

Consider two persistence points $p_0=(b_0,d_0)$ and $p_1=(b_1,d_1)$. 
The distance between $p_0$ and $p_1$ is defined by
$
\|(b_0,d_0)-(b_1,d_1)\|_\infty := \max\setof{|b_0-b_1| , |d_0-d_1|}.
$
Now, given two persistence diagrams $\pd$ and $\pd'$  let $\gamma\colon \pd \to\pd'$ be a bijection between points in the two persistence diagrams where we are allowed to match points 
of one diagram with points on the diagonal of the other diagram. The {\em degree-$q$ Wasserstein distance}, W$_q (\pd,\pd')$, is obtained by considering for each bijection, $\gamma$, the quantity
\[
\left(\sum_{p\in\pd} \| p -\gamma(p)\|^q_\infty \right)^{1/q}
\]
and defining the distance between $\pd$ and $\pd'$ to be the minimum value of this quantity over all possible bijections.
Stated formally
\[
{\rm W}_q (\pd,\pd') = \inf_{\gamma\colon \pd \to \pd'} \left(\sum_{p\in\pd} \| p -\gamma(p)\|^q_\infty \right)^{1/q}. 
\] 
In simple terms, ones looks into the cheapest way to move the points of one diagram to the those corresponding
to the other diagram, remembering that the points could be moved to the diagonal as well. 
The cost of `moving the points'  (i.e., selecting a given bijection) varies for different values of $q$. For example, 
the  Wasserstein distance W$_1$  sums up all the differences with equal weight. For $q>1$, the distance 
still keeps track of all the changes but the small differences contribute less. For brevity, in this paper we use $q = 2$ only. 
The code used for calculating distances is available online~\cite{codes}.

\section{Results}
\label{sec:results}

\subsection{Material Response: Structure and Geometry}

\begin{figure}
 \includegraphics[width=1.0\columnwidth]{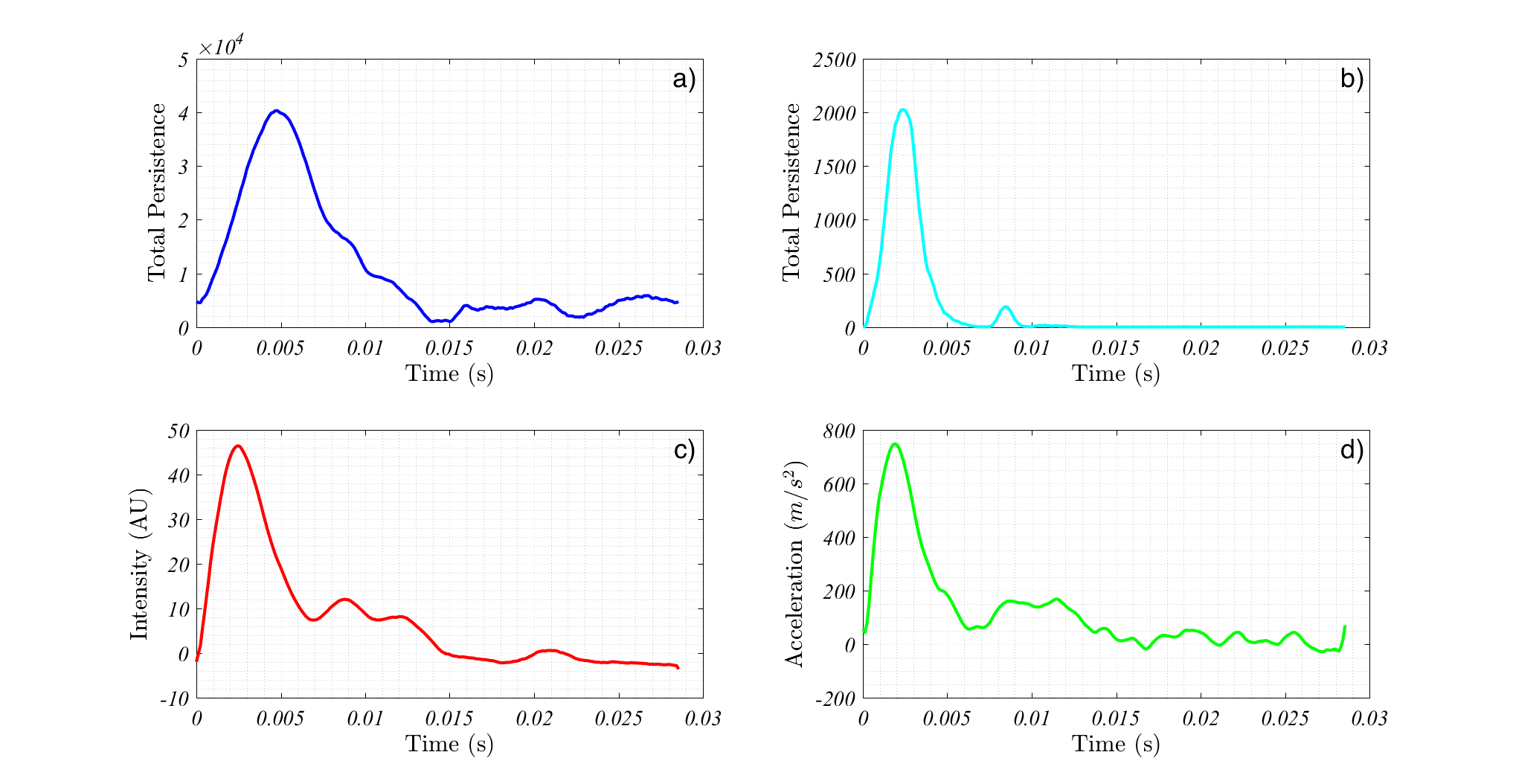}
\caption{Total persistence, for the components, TP$_0$, for the loops, TP$_1$, photoelastic response, and intruder's acceleration
 for the reference case (medium $6.35-4.9$). Supplementary animation~\cite{sup_anim_persistence} shows the above curves jointly 
with the (processed) photoelastic images. }
\label{fig:acceleration}
\end{figure}

Figure~\ref{fig:acceleration}  shows the results for the reference case. The four panels in the figure show the total persistence, TP$_0$, TP$_1$, total photoelastic intensity, and intruder's acceleration. Since the position data from which the intruder acceleration is computed are rather noisy, as discussed in~\cite{clark_prl15}, the velocity and acceleration are computed by numerical derivatives calculated based on a linear best fit to the data over a range of roughly $50$ frames. This is equivalent to taking numerical derivatives and then smoothing with a low-pass filter. For consistency, the intensity and TP data are also filtered using a similar procedure, but the photoelastic signals are less noisy and thus not as strongly affected by this process. 

We first make the general observation that TP curves are highly similar in shape to the photoelastic response and acceleration of the intruder, suggesting that the material response responsible for slowing down the intruder involves formation of highly structured force field. It is not only that the particles exposed to impact light up, but they respond in a manner that corresponds to strongly nonuniform force field. Strongest de-acceleration of the intruder (the peak in Fig.~\ref{fig:acceleration}(d)) corresponds (approximately) to the maximum values of TP$_0$ and TP$_1$. 

Proceeding with the more precise comparison of the curves, we observe particularly good agreement among TP$_1$, photoelastic response, and acceleration. Examining the acceleration curve in Fig.~\ref{fig:acceleration}(d), we note that there is a primary peak, corresponding to forces building up beneath the intruder and relaxing, but then there is a secondary rise at roughly 0.008 seconds. This event is clearly visible in movies of the impact, see~\cite{sup_anim_persistence}. TP$_1$ captures the secondary rise in the acceleration at roughly 0.008 seconds, while TP$_0$ does not follow the acceleration and photoelastic intensity accurately. This result already suggests that the properties of the loops, in particular, play an important role in determining the dynamics of the intruder. 

\begin{figure}
 \includegraphics[width=1.0\columnwidth]{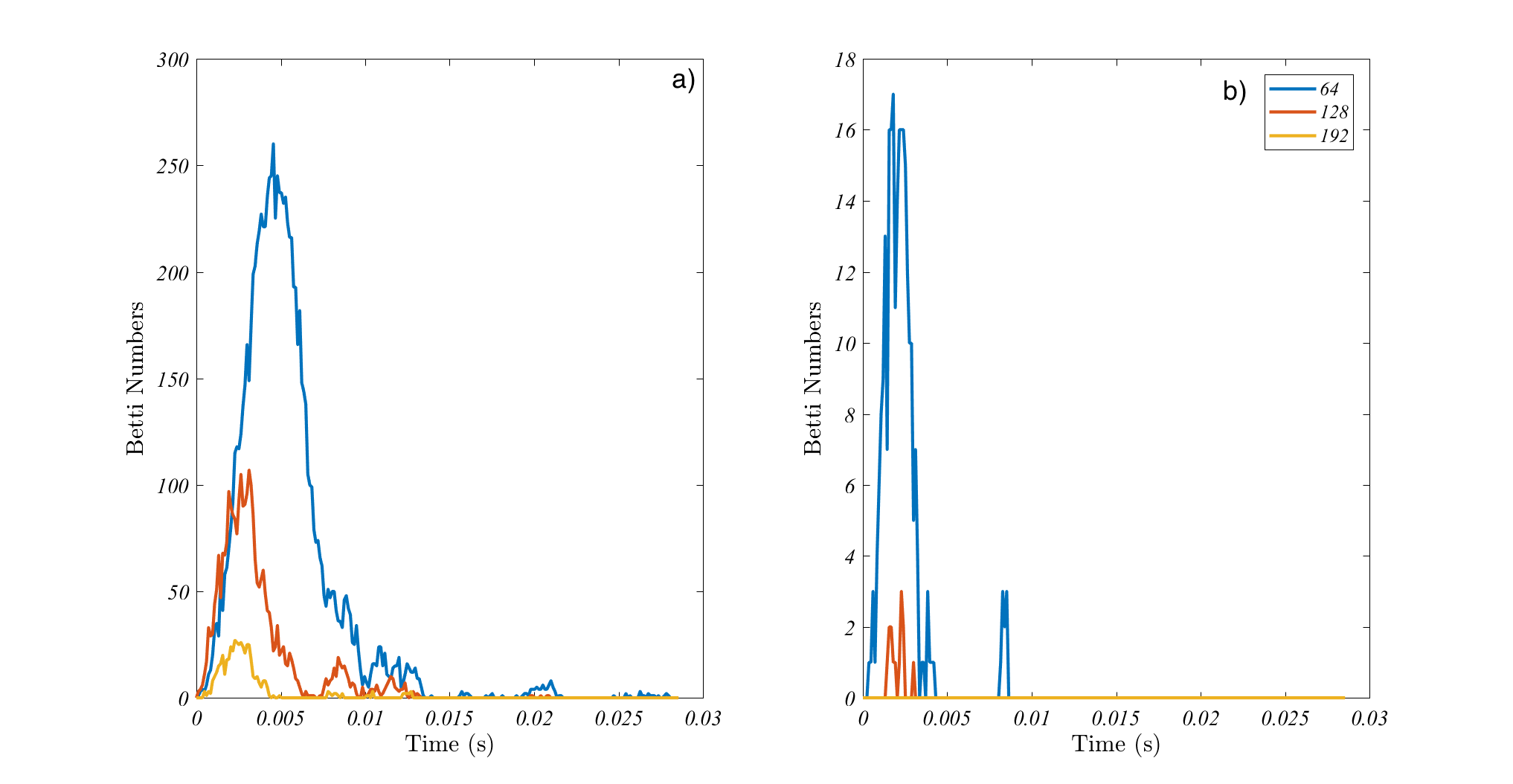}
\caption{Betti numbers, $\beta_0$ (a)  and $\beta_1$ (b),  for the reference case, respectively.  No smoothing is used here since the amount of 
data particularly for $\beta_1$, is limited. The results are shown for three different brightness threshold levels, as discussed
in the text.
 }
\label{fig:betti}
\end{figure}

Since TP provides such a good description of the dynamics of the intruder, one may wonder whether some simpler quantity could provide similarly good description. For example, one could ask whether it is indeed necessary to employ a relatively complicated concept of persistence on the first place, and instead resort to a simpler measure, such as Betti numbers. In Sec.~\ref{sec:topology} we described how to compute Betti numbers from the persistence, however Betti numbers could be computed directly as well by simply counting the number of components and loops for a given intensity threshold. Figure~\ref{fig:betti} shows the Betti numbers, $\beta_0$ and $\beta_1$ for the reference case, and for few different threshold levels. Clearly, as the threshold level is increased, the number of components and loops decreases sharply, particularly for the loops. Also, by comparison with the Fig.~\ref{fig:acceleration}, we note that the degree of agreement between the acceleration/photoelastic response for Betti numbers is much weaker. This is not surprising since  Betti numbers provide only partial information. Therefore, one cannot expect such precise correlation with the intruder's dynamics and photoelastic response as it was the case with the TP. In addition, by definition Betti numbers involve choice of threshold, therefore any conclusion obtained using Betti numbers is based on an (arbitrary) choice.

\begin{figure}
 \includegraphics[width=1.0\columnwidth]{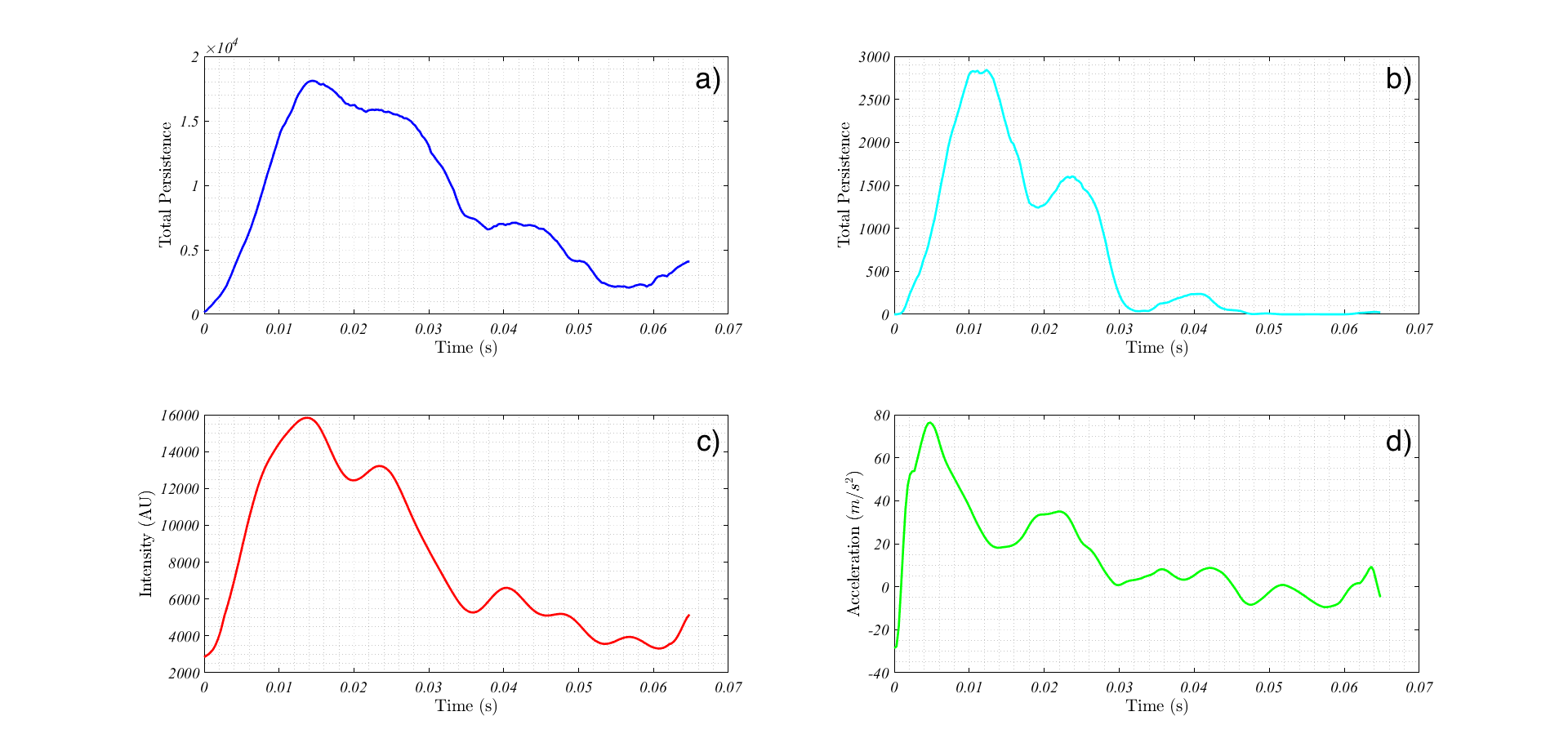}
 \caption{
 Total persistence, for the components, TP$_0$, for the loops, TP$_1$, photoelastic response, and intruder's acceleration
 for the impact on soft particles (soft $6.35-1.3$). 
 }
\label{fig:acceleration_soft}
\end{figure}

So far we have focused on a specific experiment considering impact on the particles of medium 
stiffness. Figure~\ref{fig:acceleration_soft}  shows another example of impact on soft particles. This example
is  interesting since it shows particularly clearly the correlation between the TP$_1$ and 
the acceleration of the intruder. The agreement between TP$_1$ and acceleration
is even better than the agreement between photoelastic response and acceleration. We conclude that the 
strong presence of loops in the force networks (as quantified by the TP$_1$) appears to plays an important role in determining the 
intruders dynamics. Since the times at which TP$_1$ and acceleration are significant closely coincide, we 
conjecture that the presence of the loops is important factor in slowing down of the intruder. 

Table~\ref{tab:table} shows in precise terms the degree of correlation between the various topological measures 
and the photoelastic response for the $10$ considered experiments. Across the board, we find high correlation 
for the measures related to loops, showing once again that the loops play an important role across different impact
speeds and different particle properties. 

The table also shows TC$_{0,1}$, defined as the total number of generators for the components and loops, respectively. This 
measure appears to also show similar level of correlation as TP. 

\begin{table}
\caption{The correlations between photoelastic response and various topological measures. The value of $1$ would correspond 
to perfect correlation, while $0$ would mean complete lack of correlation. 
 }
\label{tab:table}
\begin{tabular}{|c c  c c c|} 
 \hline \hline
Experiment \hspace{0.5in}    & TP$_0$ \hspace{0.5in}  & TP$_1$ \hspace{0.5in}  &   TC$_0$ \hspace{0.5in}     &   TC$_1$    \\ [0.5ex] 
 \hline
  medium $6.35-2.2$    \hspace{0.5in}    & $0.95$ \hspace{0.5in}     & $ 0.46$  \hspace{0.5in}  &     $0.76$  \hspace{0.5in}   &   $0.56$    \\  
  medium $6.35-3.2$    \hspace{0.5in}    & $0.90$ \hspace{0.5in}     & $ 0.69$  \hspace{0.5in}  &     $0.72$    \hspace{0.5in}   &   $0.79$    \\
  medium $6.35-4.9$    \hspace{0.5in}    & $0.66$ \hspace{0.5in}     & $ 0.91$  \hspace{0.5in}  &     $0.66$   \hspace{0.5in}   &   $0.94$    \\
  medium $12.7-2.6$    \hspace{0.5in}    & $0.40$ \hspace{0.5in}     & $ 0.91$  \hspace{0.5in}  &     $0.25$  \hspace{0.5in}   &   $0.91$    \\  
  medium $12.7-3.8$    \hspace{0.5in}    & $0.71$ \hspace{0.5in}     & $ 0.95$  \hspace{0.5in}  &     $0.64$  \hspace{0.5in}   &   $0.92$    \\
  medium $12.7-4.5$    \hspace{0.5in}    & $0.87$ \hspace{0.5in}     & $ 0.84$  \hspace{0.5in}  &     $0.66$   \hspace{0.5in}   &   $0.88$    \\
  ~medium $20.32-2.3$    \hspace{0.5in}    & $0.87$ \hspace{0.5in}     & $ 0.96$  \hspace{0.5in}  &     $0.84$   \hspace{0.5in}   &   $0.96$    \\
  soft $6.35-1.3$    \hspace{0.5in}     & $0.91$  \hspace{0.5in}    & $ 0.94$  \hspace{0.5in}  &     $0.83$  \hspace{0.5in}   &   $0.97$    \\   
  soft $6.35-2.3$    \hspace{0.5in}     & $0.92$ \hspace{0.5in}     & $ 0.98$  \hspace{0.5in}  &     $0.67$   \hspace{0.5in}   &   $0.96$    \\   
  soft $6.35-3.4$    \hspace{0.5in}     & $0.92$ \hspace{0.5in}     & $ 0.92$  \hspace{0.5in}  &     $0.77$   \hspace{0.5in}   &   $0.96$    \\    \hline 
  average             \hspace{0.5in}     & $0.81$ \hspace{0.5in}     & $ {\bf 0.86}$  \hspace{0.5in}  &     $0.63$   \hspace{0.5in}   &   ${\bf 0.89}$    \\    
    \hline \hline
\end{tabular}
\end{table}

\subsection{Material Response: Time Evolution of Networks}

\begin{figure}
 \includegraphics[width=0.49\columnwidth]{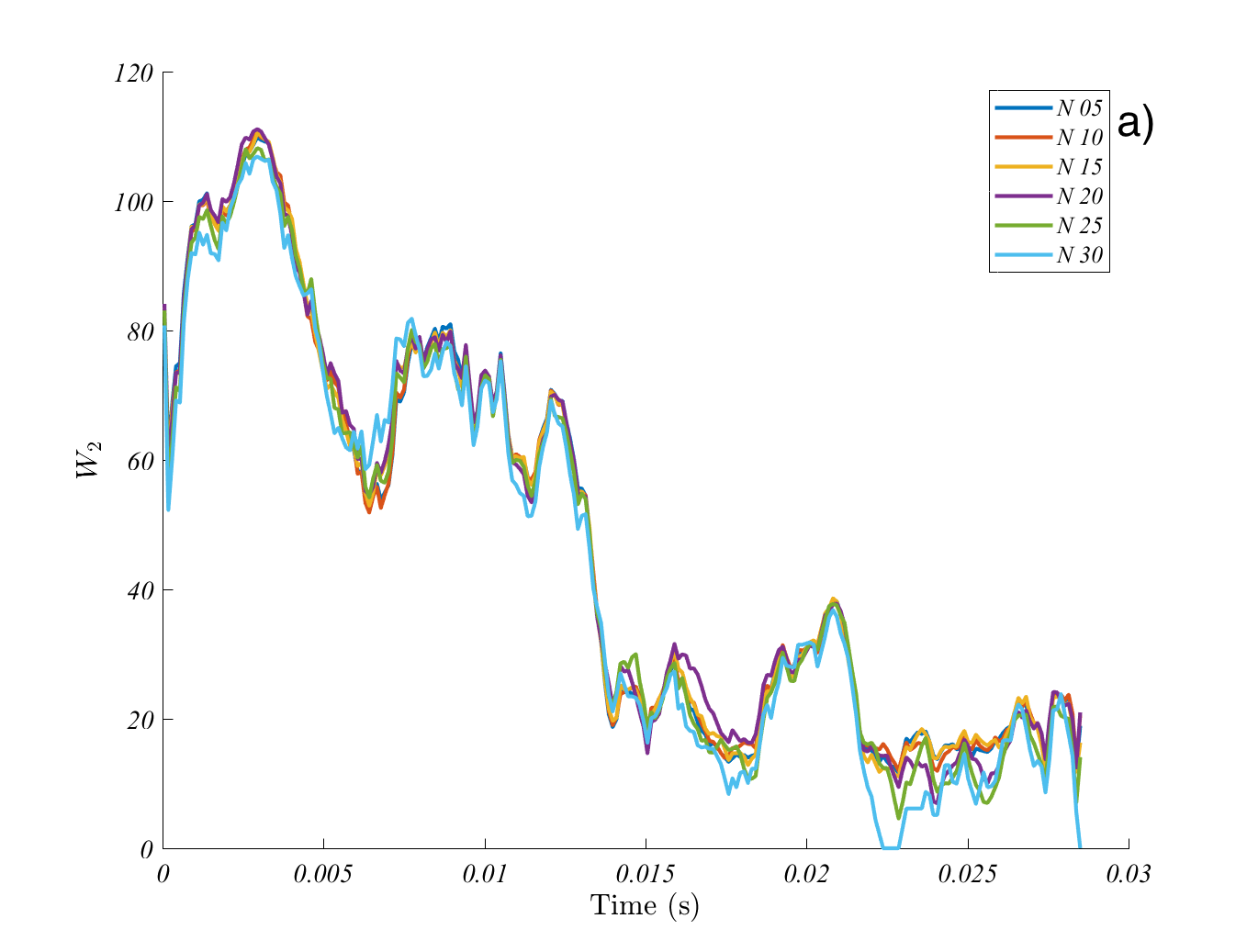}
  \includegraphics[width=0.49\columnwidth]{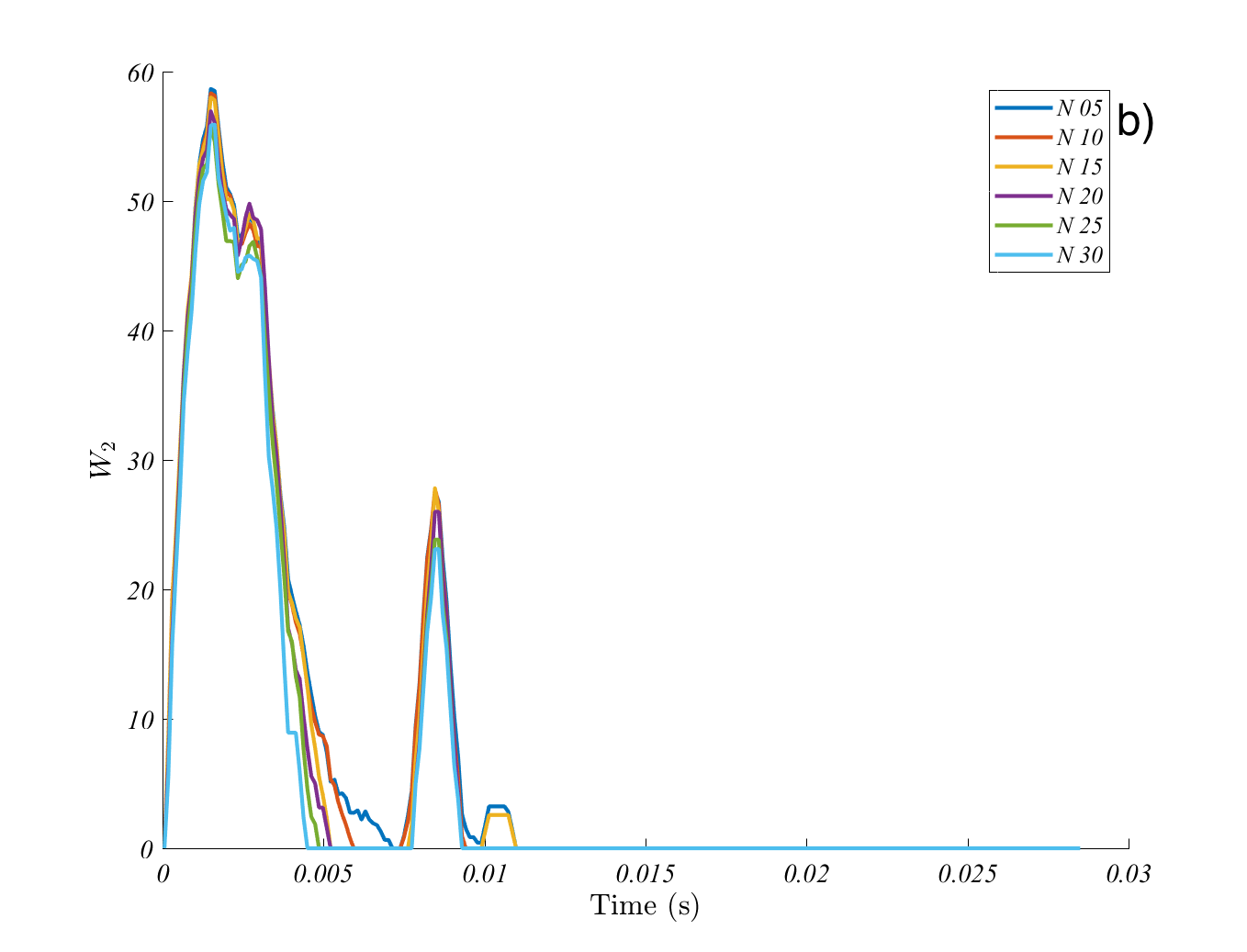} 
 \caption{Wasserstein distance between the consecutive images for the reference case as a different amount of noise (shown by the values
 of $N$ in the legends is assumed for
 components (a) and loops (b). The distances are essentially insensitive to noise. 
}
\label{fig:wnoise_ref}
\end{figure}

Persistent homology allows for extraction not only of stationary properties of force networks, but also of their evolution using 
the distance concept discussed in Sec.~\ref{sec:techniques}. 
Calculation of distance is computationally demanding, since the analysis of a single experiment involves 
comparing large number of images, and each image includes a large number of generators (the computational cost 
of carrying the computation is O$(N^3)$, where $N$ is the number of generators). 
Of course, one could skip the images to simplify the calculations (we will do this in what follows, although for a different purpose); however, 
such an approach leads to loss of information. Another approach that is more appropriate is based on the fact that the persistence diagrams 
involve large number of generators close to the diagonal, see Sec.~\ref{sec:topology} and supplementary animation~\cite{sup_anim_persistence}. 
These generators could be thought of as noise, since they represent the features that persist only for a very small range of 
the force thresholds/image brightness levels; this range may be even close to the accuracy of the process leading to the analyzed experimental images. 
Therefore, it makes sense to remove some of the generators very close to the diagonal; since there a many generators there, the 
computations that ignore these generators can be carried much faster. Figure~\ref{fig:wnoise_ref} shows the results obtained if the generators corresponding 
to the specified number of brightness levels are not considered: we see that the distance between the images is essentially insensitive if the generators
up to $30$ levels from the diagonal are removed. We therefore carry out further computations of the distance ignoring the generators that are closer
than $30$ levels to the diagonal. 

\begin{figure}
 \includegraphics[width=0.495\columnwidth]{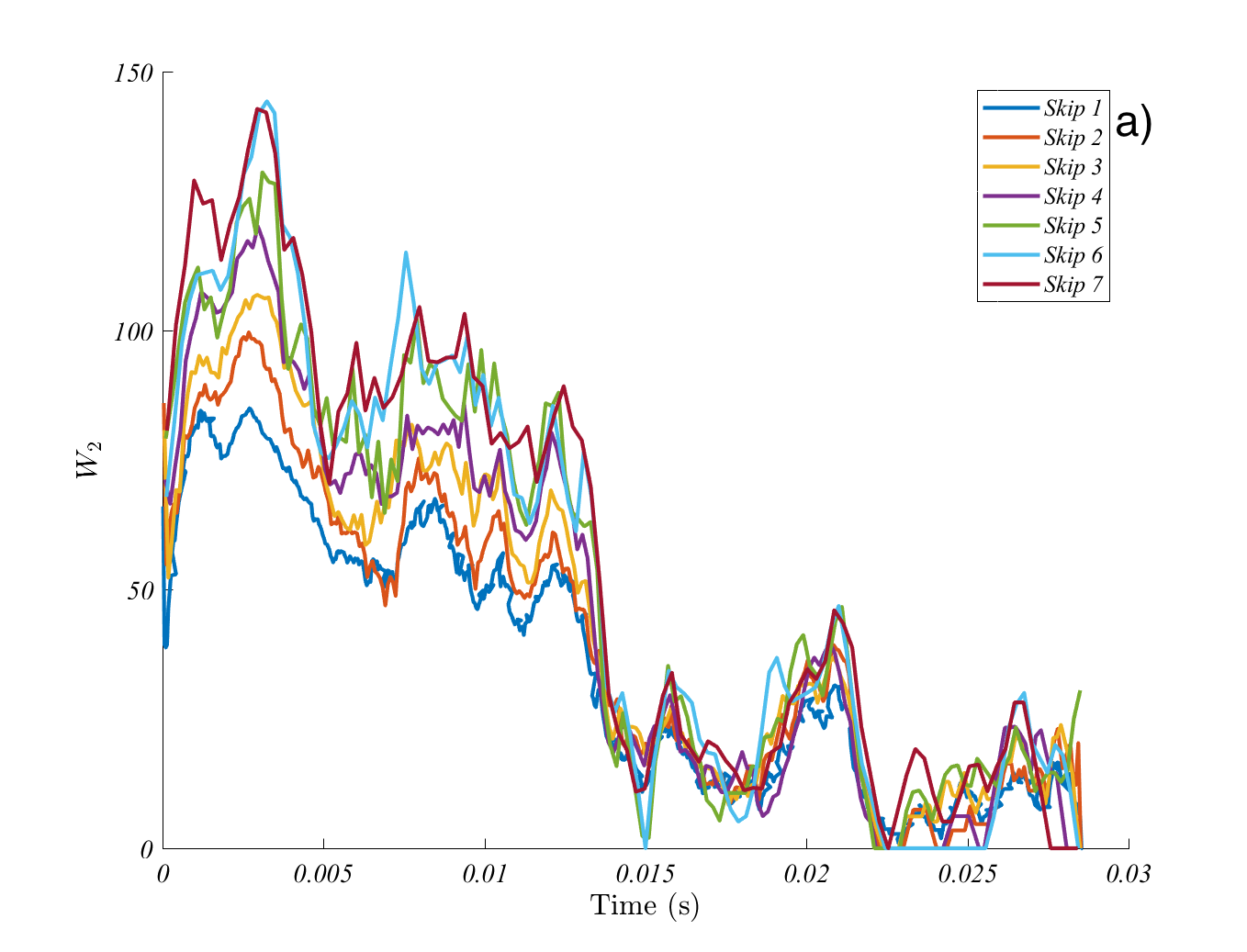}
  \includegraphics[width=0.495\columnwidth]{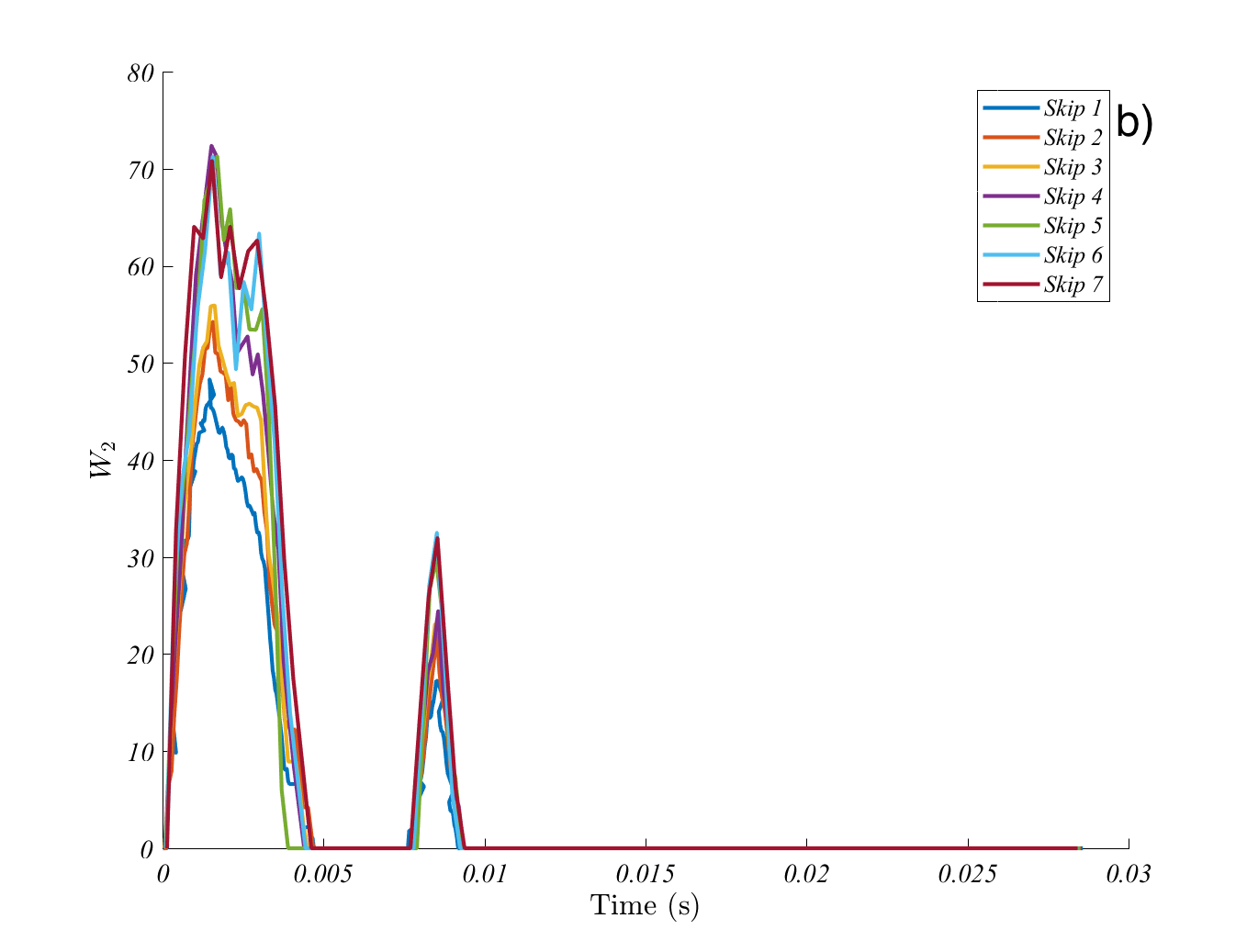} 
 \caption{Wassertein distance between the images for the reference case assuming that the generators closer than $30$ pixels to the diagonal  can be ignored. 
The given number of `skips' specifies the number of images that were skipped in the calculation of the Wasserstein distance.   The part (a) shows the results
for components, and the part (b) for loops.}
\label{fig:wskip_ref}
\end{figure}

Figure~\ref{fig:wskip_ref} shows the results obtained for the distance if the specified number of images is skipped in calculations.
The results show that, as expected, the distance between the images increases with the number of skipped images. The increase is, 
however, slow: even if only every $7$th image is considered, the distance increases by less than a factor of $2$. The significance
of this finding is as follows: if the evolution of the force field were completely resolved, the distance should have increased linearly
with the number of images skipped. This is not the case, however, showing that the evolution of the force field during impact is {\it not}
completely resolved, meaning that the force field evolves on the time scale which is faster than the inverse sampling rate used. While,
based on the results provided, we do not know what is the time scale on which the force networks evolve, at least the presented 
results provide a lower bound. The final conclusions is therefore that the force network evolve on the time scale which is not 
completely captured by the experimental imaging. We note that the same conclusions can be reached
by considering soft particles (figures not shown for brevity).

\section{Conclusions}
\label{sec:conclusions}

In this paper, we have shown the utility of persistent homology in analyzing the results of physical experiments that are 
difficult to analyze via other means. Due to finite resolution of the experimental images, and fast evolution, the quality of the 
available data is far from perfect; still, persistent homology provides meaningful and insightful results regarding the structure
of the force networks in a granular system exposed to impact. 

Persistent homology allows quantification of structural or geometrical properties of the force networks. This analysis has shown, for a set of 
$10$ considered experiments that involve different impact speeds and different particle properties, that the loop structure 
of the force networks is crucial in understanding material response and the dynamics of the intruder. It is important to 
point out that this conclusion does not involve the concept of force threshold: it extends over all force thresholds. Therefore, 
arbitrariness  associated with the choice of force threshold is avoided.   Furthermore, we note that the the utility of 
persistent homology is not limited to two spatial dimensions (2D): as long as the data are available, similar type of analysis 
can be carried out in 3D.

The tools of persistent homology also allow quantification of dynamical properties of the force networks. We are in the position to compare, in a quantitative manner, the experimental images obtained at different instances in time during the experiments. This comparison allows further to discuss the time scale on which force networks evolve. Surprisingly, we are able to show that the evolution happens on the time scale that is faster than the inverse frame rate used in the experiments. Further work will be needed to capture this time scale in more precise terms.                                                                                                                                                                                                                                                                                                                                                                                                                                                                                                                                                                                                                                                                                                                                                                                                                                                                                                                                                                                                                                                                                                                                                                                                                                                                                                                                                                                                                                                                                                                             

\begin{acknowledgments}
The experiments discussed in the present paper were carried out in the laboratory of Robert P. Behringer at Duke University. The authors also acknowledge many useful discussions with Lenka Kovalcinova, as well as with the 
members of the computational topology group at Rutgers University, in particular, Rachel Levanger, Miroslav Kram\'ar, and Konstantin Mischaikow. This work was supported in part by the NSF grant  
No. DMS-1521717 and DARPA grant HR0011-16-2-0033. 
\end{acknowledgments}


%

\end{document}